\documentclass[preprint,amssymb]{revtex4}
\usepackage{graphicx}

\begin{document}

\title{Macroscopic Quantum Tunnelling in Rotating Bose-Einstein Condensates}
\author{ Yuan-Xiu Miao, Hui Zhai, Hong-Yu Yang and Lee Chang\\ Center for Advanced Study,
Tsinghua University, 100084\\Beijing, China}
\date{\today}
\begin{abstract}
In this paper we investigate the macroscopic quantum tunnelling
and the phase coherence property of the rotating Bose-Einstein
condensates in both static and dynamic cases by using the mean
field theory.
\end{abstract}
\maketitle

\section{Introduction}
It is well known that the phase coherence and the macroscopic
quantum phenomenon are among the important characteristics of the
Bose-Einstein condensates ( BEC ), and have induced a lot of
theoretical and experimental interests. Now, we came to know that
the phase coherence property and the macroscopic tunnelling are so
closely related that such a system was found to exhibit quite rich
physics. On the other hand, motivated by the superfluidity of the
dilute Bose gas, much work are focused on the rotating BEC
recently \cite{Leggett}. In this paper, we investigate the
macroscopic quantum tunnelling in the rotating BEC in various
parameter ranges and in both static and dynamic cases with the
help of the mean field approach, and the phase coherence property
in this system will be emphasized.

In Section~\ref{sec:2}, a general description of our system and
the Hamiltonian to begin with are given. In the next two sections,
we focus on the static property when the rotating frequency is
fixed. SU(2) coherent state path integral method is used in
Section~\ref{sec:3} and we find that if the parameters satisfy a
specific condition, the ground state represents the superposition
of two macroscopic quantum states. The energy splitting due to the
macroscopic coherent tunnelling is also calculated by the
instanton technique. In another parameter range, fluctuations of
the phase and the particle number of the ground state are given in
Section~\ref{sec:4} and it follows that the condition of the
Josephson effect in this case is satisfied. In
Section~\ref{sec:5}, we consider the case where the rotating
frequency varies with time, and the nonlinear Landau-Zener
tunnelling \cite{QNiu} is realized in this system. Finally, some
conclusions and further discussion are given in
Section~\ref{sec:6}.

\section{\label{sec:2} Model}
We consider the Bose-Einstein condensate with total particle
number $N$ which is trapped in a circularly symmetrical trap $V$
or container, and rotated by an asymmetric perturbation $V_{ext}$.
The N-body Hamiltonian in the rotating frame of such a system is
given by $H=\sum
\limits_{n=1}^{N}[H_{0n}+V_{ext}(\vec{r}_n)-\omega \cdot
L_{zn}]+\sum\limits_{1\leq i<j\leq N}U\delta
(\vec{r}_{i}-\vec{r}_{j})$, where $H_{0n}=\vec{p}_n^{\; 2
}/2m+V(\vec{r}_n)$ is the single body Hamiltonian in the
laboratory frame, $U=4\pi a \hbar^2/m$, $a$ is the s-wave
scattering length and $a > 0$ for repulsive interactions, $\omega$
is the angular frequency of the rotating frame. The lowest-lying S
and P states of the Hamiltonian $H_{0}$ are separated by an energy
$\hbar\omega_{0}$. Following \cite{Leggett}, we adopt the two-mode
approximation, i.e. under the condition that the effective
interacting energy $Un$ is much smaller than $\hbar\omega_0$ and
the energy of the first radial excitation is much large than
$\hbar\omega_0$, where $n$ is the particle number density, we
expand the field operator in terms of the single particle wave
functions of the two lowest-lying states which are exact
eigenstates of $L_z$, $\Psi (\vec{r})=a_{1}\varphi
_{1}(\vec{r})+a_{0}\varphi _{0}(\vec{r})$. Following the
conventional procedure and omitting constant terms proportional to
$N$, we get the second quantization Hamiltonian
\begin{equation}
H=\hbar
\delta\omega(a_0^{+}a_0-a_1^{+}a_1)-V_0(a_0^{+}a_1+a_1^{+}a_0)+\frac{U}{2}
(\tilde{g}_0a_0^{+}a_0^{+}a_0a_0+\tilde{g}_1a_1^{+}a_1^{+}a_1a_1+
4\tilde{g}_{01}a_0^{+}a_0a_1^{+}a_1) \label{eq:three},
\end{equation}
where $\delta\omega=\frac{1}{2}(\omega-\omega_0$), the coupling
energy $V_0 \equiv
-\int\varphi_1^*(\vec{r})V_{ext}\varphi_0(\vec{r}) d^3r$ is set
real and positive by fixing a proper phase difference between
$\varphi_0$ and $\varphi_1$ and it can be adjusted by changing the
strength of the perturbation. The $\tilde{g}$'s are determined by
the eigenfunctions $\varphi_{0}$ and $\varphi_{1}$ in the
following way $\tilde{g}_{0}\equiv \int |\varphi_0(\vec{r})|^4
d^3r$, $\tilde{g}_{1}\equiv \int |\varphi_1(\vec{r})|^4 d^3r$,
$\tilde{g}_{01}\equiv \int
|\varphi_0(\vec{r})|^2|\varphi_1(\vec{r})|^2 d^3r$.

\section{\label{sec:3} Macroscopic Quantum Coherent Tunnelling }
With the help of the mapping
\begin{eqnarray}
  \frac{1}{2}(a_0^+a_0-a_1^+a_1) & \rightarrow & J_z ,\nonumber\\
  \frac{1}{2}(a_0^+a_1+a_1^+a_0) & \rightarrow & J_x ,\nonumber\\
-\frac{1}{2}i(a_0^+a_1-a_1^+a_0) & \rightarrow & J_y
,\nonumber\\
  \frac{1}{2}(a_0^+a_0+a_1^+a_1) & \rightarrow & J ,
\end{eqnarray}
our problem is reduced to the one of a particle with large spin
$J$ in an arbitrary $\vec{J}$-dependent potential, i.e. the
Hamiltonian of the system can be rewritten as when dropping lower
terms in self-interactions
\begin{equation}
H=\hbar\delta\omega
J_z-V_0J_x+\frac{U}{2}[\tilde{g}_0(J+J_z)^2+\tilde{g}_1
(J-J_z)^2+4\tilde{g}_{01}(J^2-J_z^2)]
\end{equation}

First of all, we introduce the SU(2) coherent states
\cite{WMZhang}
\begin{eqnarray}
|\Omega\rangle & = & |\theta,\phi\rangle=e^{(\zeta J_{+}-\zeta^{*}
J_{-})}
|\frac{N}{2},-\frac{N}{2}\rangle \nonumber\\
               & = & \sum_{n=-\frac{N}{2}}^{\frac{N}{2}}\sqrt{\frac{N!}{(\frac{N}{2}+n)!
(\frac{N}{2}-n)!}}(\cos\frac{\theta}{2})^{\frac{N}{2}-n}[\exp(-i\phi)\sin
\frac{\theta}{2}]^{\frac{N}{2}+n}|n\rangle
\label{eq:one}.
\end{eqnarray}
Here,
\begin{equation}
\zeta=\frac{\theta}{2}e^{-i\phi} \; \; \; \;  (0\leq \theta \leq
\pi, 0\leq \phi \leq 2\pi) ,
\end{equation}
$|\frac{N}{2},-\frac{N}{2}\rangle$ is the extremal state
$|J=\frac{N}{2},J_z=-\frac{N}{2}\rangle$, $|n\rangle$ is the
eigenstate of $J_z$, $2n$ represents the occupation number
difference between the two states and $\phi$ stands for the phase
difference between them which will be given explicitly in the next
section. SU(2) coherent states are overcomplete in the Hilbert
space and any wave function can be expanded in terms of them, for
example, a Fock state $|n\rangle$ can be written as
\begin{equation}
|n\rangle\propto\int_{0}^{2\pi}d\phi \;
e^{i(n+N/2)\phi}|\Omega\rangle \label{eq:four}.
\end{equation}

In the SU(2)-coherent-state path-integral representation, the
Euclidean transition amplitude from an initial state to a final
state is \cite{WMZhang}
\begin{equation}
\langle
\Omega_f\left|e^{-H(\tau_f-\tau_i)/\hbar}\right|\Omega_i\rangle
=\int[d\Omega(\tau)]\exp\left[-\frac{1}{\hbar}S_E(\theta,\phi)\right]
,
\end{equation}
where
\begin{equation}
S_E(\theta,\phi)=\int_{\tau_i}^{\tau_f}d\tau\left[i\hbar\frac{N}{2}
(1-\cos\theta)\left(\frac{d\theta}{d\tau}\right)+E(\theta,\phi)\right]
\label{eq:two}.
\end{equation}
The first term in Eq.~(\ref{eq:two}) is the Wess-Zumino term, and
the semiclassical energy $E(\theta,\phi)$ accurate to a constant
proportional to $N$ is given by
\begin{equation}
\frac{1}{N}E(\theta,\phi)=[\hbar\delta\omega+\frac{1}{4}(g_0-g_1)]\cos\theta
-V_0\sin\theta\cos\phi+\frac{1}{2}[g_{01}-\frac{1}{4}(g_0+g_1)]\sin^2\theta
,
\end{equation}
where
\begin{eqnarray}
   g_0 & = & U\tilde{g}_0N,\nonumber\\
   g_1 & = & U\tilde{g}_1N,\nonumber\\
g_{01} & = & U\tilde{g}_{01}N.
\end{eqnarray}
Since $\sin\theta \geq 0$ $(0\leq \theta \leq \pi)$, for $V_0>0$
the minima of the semiclassical energy can only occur for
$\phi=0$.

It turns out that the qualitative behavior of $E(\theta,\phi)$
depends on a relation between $V_0$ and the $g$'s. The first
regime is characterized by the presence of two minima. For
$\delta\omega\neq \frac{1}{4\hbar}(g_1-g_0)$ the minima define a
ground state and a metastable state. For $\delta\omega
=\frac{1}{4\hbar}(g_1-g_0)$ the two minima are degenerate, and a
situation of macroscopic quantum coherence occurs. In the other
regime $E$ as a function of $\theta$ has one minimum, the position
of which depends on $\delta\omega$. The ground state is a coherent
superposition of $\varphi _{0}$ and $\varphi _{1}$. In this
section, we confine our discussion in the first regime.

For $g_{01}-\frac{1}{4}(g_0+g_1)>0$ and
$V_0<g_{01}-\frac{1}{4}(g_0+g_1)$, there always exists
$\delta\omega=\frac{1}{4\hbar}(g_1-g_0)$, such that there exist
two degenerate minima in the semiclassical energy. In this case,
the true ground state of the system will be the superposition of
the two macroscopic states due to the coherent quantum tunnelling
between them which may be observed by interference experiments in
principle.

As we have calculated explicitly, such conditions can be satisfied
by a large class of systems such as Bose gas in an annulus or a
thin cylindrical container. Tunnelling between the two degenerate
macroscopic states leads to energy splitting between them. We
evaluate this tunnelling splitting by applying the dilute
instanton gas approximation which is valid for large-N.

By adding some constants, the semiclassical energy can be
rewritten as
\begin{equation}
E(\theta,\phi)=Ng(\sin\theta-\sin\theta_0)^2+NV_0\sin\theta(1-\cos\phi),
\end{equation}
where $g\equiv \frac{1}{2}[g_{01}-\frac{1}{4}(g_0+g_1)]$,
$\sin\theta_0=V_0/2g$. From $\delta S_E(\theta,\phi)=0$, we get
the instanton solution corresponding to the transition from
$\theta=\theta_0$ to $\theta=\pi-\theta_0$:
\begin{eqnarray}
\bar{\theta} & = & \arccos[-\cos\theta_0\tanh(\omega_b\tau)]\nonumber\\
\bar{\phi}   & = & \arcsin\left[-\frac{i}{2}\frac{\cot^2\theta_0
\textnormal{sech}^2(\omega_b\tau)} {[1+\cot^2\theta_0
\textnormal{sech}^2(\omega_b\tau)]^{1/2}}\right]
\end{eqnarray}
where $\omega_b=2g\cos\theta_0/\hbar$. The associated instanton
action is
\begin{equation}
S_{inst}=N\left[-\cos\theta_0+\frac{1}{2}\ln\left(\frac{1+\cos\theta_0}
{1-\cos\theta_0}\right)\right],
\end{equation}
and the tunnelling splitting is
\begin{equation}
\triangle
E=32g\left(\frac{N}{2\pi}\right)^{\frac{1}{2}}\frac{(\cos\theta_0)^{5/2}}{\sin\theta_0}
\left(\frac{1-\cos\theta_0}{1+\cos\theta_0}\right)^{\frac{1}{2}\cos\theta_0}e^{-S_{inst}}.
\end{equation}

\section{\label{sec:4} Josephson tunnelling}
For $V_0>g_{01}-\frac{1}{4}(g_0+g_1)$ and arbitrary
$\delta\omega$, there always exists only one minimum in the
semiclassical energy. In this section we evaluate the fluctuations
for the relative number and phase around the ground state firstly.
In the case where $V_0>g_{01}-\frac{1}{4}(g_0+g_1)$ and
$\delta\omega=\frac{1}{4\hbar}(g_1-g_0)$, the classical ground
state of the system is $\bar{\theta}=\pi/2$, $\bar{\phi}=0$.
Expanding the Euclidean action $S_E(\theta,\phi)$ to the
second-order around its classical value $\delta S_{cl}=0$ and
performing integrations over
$\phi_1(\tau)=\phi(\tau)-\bar{\phi}(\tau)$ and
$\theta_1(\tau)=\theta(\tau)-\bar{\theta}(\tau)$ respectively
\cite{Zhou}, we get the effective actions for $\theta_1$ and
$\phi_1$ respectively
\begin{eqnarray}
I(\theta_1) & = & \int_{\tau_i}^{\tau_f} \left[
\frac{\hbar^2N}{8V_0}\dot{\theta}^2_1+\frac{N}{2}(V_0+2g)\theta_1^2
\right] \; d\tau
\;,\nonumber\\
  I(\phi_1) & = & \int_{\tau_i}^{\tau_f} \left[
  \frac{\hbar^2N}{8(V_0+2g)}\dot{\phi}_1^2+\frac{V_0N}{2}\phi^2_1
  \right] \; d\tau \;.
\end{eqnarray}
The motions of $\theta_1$ and $\phi_1$ are approximately harmonic
oscillators and we get the relative number fluctuation
\begin{equation}
\triangle(N_0-N_1)=\sqrt{N\sqrt{\frac{V_0}{V_0-2g}}},
\end{equation}
and the relative phase fluctuation
\begin{equation}
\triangle{\phi}=\sqrt{\frac{1}{N}\sqrt{\frac{V_0-2g}{V_0}}}.
\end{equation}

In order to show the relationship between the N-body identical
particle state and the condensate wave function, we rewrite the
SU(2) coherent state $|\theta,\phi\rangle$ as
\begin{equation}
|\theta,\phi\rangle=\frac{1}{\sqrt{N!}}(e^{-i\phi}\sin{\frac{\theta}{2}}a_1^+
+\cos{\frac{\theta}{2}}a_0^+)^N|0\rangle.
\end{equation}
In coordinate representation, it corresponds to
\begin{equation}
\psi(\vec{r}_1,\cdot\cdot\cdot\vec{r}_{N})=\prod_{r_i}\left(e^{-i\phi}
\sin{\frac{\theta}{2}}\varphi_1(\vec{r}_i)+\cos{\frac{\theta}{2}}
\varphi_0(\vec{r}_i)\right).
\end{equation}
From the underlying concept of the off-diagonal-long-range-order
(ODLRO) of C.N. Yang \cite{Yang}, we have
\begin{equation}
\int d^3r_2 \cdot\cdot\cdot d^3r_N \psi^*(\vec{r},\vec{r}_2,
\cdot\cdot\cdot\vec{r}_N)\psi(\vec{r}^{\;\prime}, \vec{r}_2,
\cdot\cdot\cdot \vec{r}_N)=\frac{1}{N}\Phi^*(\vec{r})
\Phi(\vec{r}^{\;\prime}) ,
\end{equation}
and $\Phi(\vec{r})=\sqrt{N}
[e^{-i\phi}\sin{\frac{\theta}{2}}\varphi_1
+\cos{\frac{\theta}{2}}\varphi_0]$ is called the condensate wave
function.

As we have calculated, for $V_0$ sufficiently larger than $2g$ we
have $\triangle(N_0-N_1)\simeq\sqrt{N} \ll N$ and
$\triangle\phi\ll 1$. This means that $\varphi_1$ and $\varphi_0$
can be viewed as almost two condensates and their phase difference
is almost fixed for a large $N$. Thus, the ground state of the
system satisfies the condition of the Josephson effect. It's easy
to transfer the Hamiltonian (\ref{eq:three}) into a Josephson
pendulum Hamiltonian
\begin{equation}
H=\frac{4g}{N}\frac{\partial^2}{\partial\phi^2}-V_0N\sqrt{1-\cos^2\bar{\theta}}
\cos{\phi},
\end{equation}
with the Josephson frequency
$\sqrt{8gV_0\sqrt{1-\cos^2{\bar{\theta}}}}$.

It should be noted that, in the above we consider the case of
$\delta\omega=\frac{1}{4\hbar}(g_1-g_0)$ in which the
semi-classical ground state gives $\cos{\bar{\theta}}=0$. The
discussions above can be also applied to other cases in which
$\cos{\bar{\theta}}\neq 0$ and $\cos{\bar{\theta}}$ changes
gradually as $\delta\omega-\frac{1}{4\hbar}(g_1-g_0)$ increases
until its absolute value reaches $1$.

\section{\label{sec:5} Nonlinear Landau-Zener tunnelling}

Consider a two-level system, its adiabatic eigenenergies
$\varepsilon_0(t)$ and $\varepsilon_1(t)$ increases and decreases
linearly with time $t$ respectively, crossing at a certain value
of $t_0$. Taking account of the coupling energy $V_0$ between the
two levels, the adiabatic eigenvalues of such a system $E_0(t)$
and $E_1(t)$ with wavefunctions $\psi_0(t)$ and $\psi_1(t)$ avoid
crossing. We know from the adiabatic theorem that if the system
with coupling is initially in the state $\psi_0(t\ll t_0)$ and $t$
changes infinitely slowly from $t\ll t_0$ to $t\gg t_0$, then the
system will remain in the state $\psi_0(t\gg t_0)$. However, if
$t$ changes with a finite velocity, the final state of the system
will be a linear combination of $\psi_0(t\gg t_0)$ and
$\psi_1(t\gg t_0)$. This means when the two eigenenergies are
getting close to each other, the system will undergo an
instanton-like tunnelling, which is well known as Landau-Zener
tunnelling \cite{LandauZener} and can be realized in a lot of
systems in molecular and atomic physics \cite{exam}. The essential
idea of the Landau-Zener tunnelling can be generalized to the
many-body mean-field problem with non-linear interactions
\cite{QNiu}, in which the tunnelling can take place even in the
adiabatic case.

In the two previous sections we study the case where the angular
frequency $\omega$ is fixed while it can be dependent of time in
real life. Thus, it is necessary to consider the case where
$\omega$ varies with time, accompanied by a non-linear
Landau-Zener tunnelling.

The equations of motion of the operators $a_0$ and $a_1$ can be
derived from the Hamiltonian~(\ref{eq:three})
\begin{eqnarray}
i\hbar\frac{da_0}{dt} & = &
[\hbar\delta\omega+U(\tilde{g}_0a_0^+a_0
+2\tilde{g}_{01}a_1^+a_1)]a_0-V_0a_1\nonumber\;\\
i\hbar\frac{da_1}{dt} & = &
[-\hbar\delta\omega+U(\tilde{g}_1a_1^+a_1
+2\tilde{g}_{01}a_0^+a_0)]a_1-V_0a_0\;
\end{eqnarray}
Applying the mean-field approximation while dropping the
fluctuations of the field operators
\begin{eqnarray}
\langle a_1 \rangle & = & \langle a_1^+ \rangle^*  =  x_1
\sqrt{N}\nonumber\\
\langle a_0 \rangle & = & \langle a_0^+ \rangle^*  =  x_0
\sqrt{N}\;,
\end{eqnarray}
we get a non-linear Landau-Zener equation
\begin{equation}
i\hbar\frac{d}{dt} \left( \begin{array}{cc} x_0 \\ x_1 \end{array}
\right) = \left( \begin{array}{cc}
\hbar\delta\omega+d_1+c_1(|x_1|^2 -|x_0|^2) & -V_0 \\ -V_0 &
-\hbar\delta\omega+d_2-c_2(|x_1|^2
-|x_0|^2)\end{array} \right) \left( \begin{array}{cc} x_0 \\
x_1\end{array}\right) \label{eq:five}
\end{equation}
where
\begin{eqnarray}
d_1 & = & \frac{1}{2}(g_0+2g_{01})\;,\nonumber\\
c_1 & = & \frac{1}{2}(g_0-2g_{01})\;,\nonumber\\
d_2 & = & \frac{1}{2}(g_1+2g_{01})\;,\nonumber\\
c_2 & = & \frac{1}{2}(g_1-2g_{01})\;,
\end{eqnarray}
$|x_0|^2$, $|x_1|^2$ are the probabilities in corresponding states
and $|x_0|^2+|x_1|^2=1$. Our model is a more general one than that
of Ref. \cite{QNiu} for the influence of the interaction on these
two modes is different. If $\hbar\delta\omega$ varies linearly
with time, i.e. $\hbar\delta\omega=\alpha t$, we know from the
linear Landau-Zener effect that an instanton-like transition will
occur at times near to the crossing point of the unperturbed
energy levels. As the role of the parameters $d_1$ and $d_2$ in
Eq.~(\ref{eq:five}) is to shift the crossing point of the
unperturbed energy levels, we have reason to expect the
dimensionless parameter $G\equiv (g_0-g_1)/g_0$ will only affect
the transition time. Another characteristic dimensionless
parameter of the system $K\equiv (g_1-2g_{01})/(g_0-2g_{01})$ is
given by the imbalance of the particle interactions.

\begin{figure}[htbp]
\begin{center}
\includegraphics[width=3.6in]
{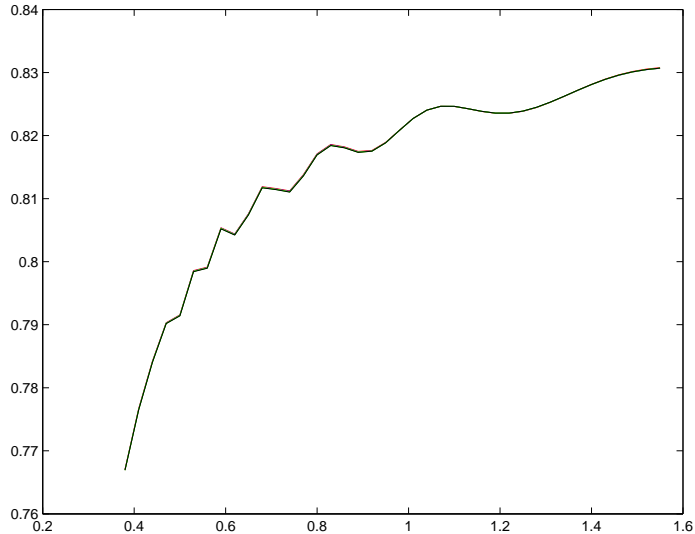} \caption{\label{fig:fig}Tunnelling probability as a
function of the dimensionless parameter $K$ is plotted for
different values ($0.1$, $0.2$, $0.3$, $0.4$, $0.5$) of the
parameter $G$. The plots for different values of the parameter $G$
are almost identical to one another.}
\end{center}
\end{figure}

The system of non-linear equations is solved numerically.
Fig.~\ref{fig:fig} shows that the transition probability as a
function of $K$ for $t\rightarrow +\infty$ is nearly monotonic
increasing. The plots with different values of the parameter $G$
are almost identical, which implies the tunnelling probabilities
are not sensitive to $G$ as we have predicted. As we can see, the
system is in a coherent state of the two states $\varphi_0$ and
$\varphi_1$ because of the nonzero transition probabilities even
in the adiabatic case. This can also explain the hysteresis
phenomenon of the vortex formation.

\section{\label{sec:6} Summary and Conclusions}
In summary, we have studied the macroscopic phenomena in rotating
BEC by using the SU(2) coherent state path-integral method. Our
discussion can also be extended to the rotating BEC with
attractive interactions and BEC in an optical lattice. In the
latter case, we can expand the field operator in terms of
single-particle ground state wave functions localized in each
well, then the coupling energy $V_0$ is replaced by the hopping
energy, which depends on the overlap of the local wave functions.
Some detailed calculation can be found in an earlier paper
\cite{Zhou}. We can see that when $V_0$ is sufficiently small, the
phase fluctuation will be very large, $\sim 2\pi$. According to
Eq.~(\ref{eq:four}), the system is in a Fock region. It can be
easily understood physically in the rotating case that when the
coupling energy is absent, the tunnelling is forbidden by the
angular momentum conservation law and the relative number
fluctuation is suppressed. In an optical lattice, small $V_0$
means sufficiently high barrier which also prohibits the
tunnelling. When $V_0$ is comparable to the self-interaction
energy, the system is in the Josephson region as we have shown in
section IV. Thus, continuously decreasing $V_0$ will lead the
system to undergo a phase-diffusion process and there is a
crossover from the Josephson to Fock region as exhibited
experimentally recently in Ref. \cite{Greiner}. The physics of
such experiments can be understood clearly and quantitatively in
our SU(2) coherent state frame.

\textit{Acknowledgements}: We are grateful to Prof. Qian Niu for
bringing Ref~\cite{QNiu} to our attentions and helpful discussion.
We also acknowledge useful discussions with Yi Zhou, Wei-qiang
Chen, Rong Lu, Duan-lu Zhou and Bei Zeng. This work is supported
by National Natural Science Foundation of China.


\begin{thebibliography}{99}

\bibitem{Leggett}
Anthony J. Leggett, Rev. Mod. Phys. {\bf 73},  307 (2001) and
references therein.

\bibitem{QNiu}
Biao Wu and Qian Niu, Phys. Rev. A {\bf 61}, 023402 (2000).

\bibitem{WMZhang}
Wei-Min Zhang, Da Hsuan  Feng and Robert Gilmore, Rev. Mod. Phys.
{\bf 62}, 867 (1990).

\bibitem{Zhou}
Yi Zhou, Hui Zhai, Rong Lu, Zhan Xu, Lee Chang, cond-mat/0104361
.

\bibitem{Yang}
C.N.Yang, Rev. Mod. Phys. {\bf 34}, 4 (1962).

\bibitem{LandauZener}
L.D. Landau, Phys. Z. Sowjetunion {\bf 2}, 46 (1932); G. Zener,
Proc. R. Soc. London, Ser. A {\bf 137}, 696 (1932).

\bibitem{exam}
W. Wernsdorfer and R. Sessoli, Science {\bf 284}, 134 (1999); K.
Mullen {\it et al.}, Phys. Rev. Lett. {\bf 60}, 1097 (1988); C.F.
Bharucha {\it et al.}, Phys. Rev. A {\bf 55}, R857 (1997); A.
Sibille {\it et al.}, Phys. Rev. Lett. {\bf 80}, 4506 (1998).

\bibitem{Greiner}
Markus Greiner, Olaf Mandel, Tilman Esslinger, Theodor W.
H$\ddot{\textnormal{a}}$nsch and Immanuel Bloch, Nature {\bf 415},
39 (2002).

\end{thebibliography}
\end{document}